\documentclass[useAMS,usenatbib,usegraphicx]{mn2e}

\title[Cir X-1 jet powered nebula]{The large-scale jet-powered radio
nebula of Circinus X-1} \author[V. Tudose et
al.]{V. Tudose,$^{1,2}$\thanks{E-mail: vtudose@science.uva.nl (VT)}
R.P. Fender,$^{3,1}$ C.R. Kaiser,$^{3}$ A.K. Tzioumis,$^{4}$ M. van
der Klis$^{1}$ \newauthor and R. Spencer$^5$\\ $^{1}$"Anton Pannekoek"
Astronomical Institute, University of Amsterdam, Kruislaan 403, 1098
SJ Amsterdam, The Netherlands\\ $^{2}$Astronomical Institute of the
Romanian Academy, Cutitul de Argint 5, RO-040557 Bucharest, Romania\\
$^{3}$School of Physics and Astronomy, University of Southampton,
Highfield, SO17 1BJ Southampton, UK\\ $^{4}$Australia Telescope
National Facility, CSIRO, PO Box 76, Epping New South Wales 1710,
Australia \\ $^{5}$University of Manchester, Nuffield Radio Astronomy
Laboratories, Jodrell Bank, SK11 9DL, Cheshire, UK}

\begin{document}

\date{Accepted XXXXXX. Received XXXXXX; in original form XXXXXX}

\pagerange{\pageref{firstpage}--\pageref{lastpage}} \pubyear{2005}

\maketitle

\label{firstpage}

\begin{abstract}

We present multi-epoch observations of the radio nebula around the
neutron star X-ray binary Circinus X-1 made at 1.4 and 2.5 GHz with
the Australia Telescope Compact Array between October 2000 and
September 2004. The nebula can be seen as a result of the interaction
between the jet from the system and the interstellar medium and it is
likely that we are actually looking toward the central X-ray binary
system through the jet-powered radio lobe. The study of the nebula
thus offers a unique opportunity to estimate for the first time using
calorimetry the energetics of a jet from an object clearly identified
as a neutron star. An extensive discussion on the energetics of the
complex is presented: a first approach is based on the minimum energy
estimation, while a second one employs a self-similar model of the
interaction between the jets and the surrounding medium.  The results
suggest an age for the nebula of $\leq 10^5$ years and a corresponding
time-averaged jet power $\geq 10^{35} \mbox{ erg s$^{-1}$}$.  During
periodic flaring episodes, the instantaneous jet power may reach
values of similar magnitude to the X-ray luminosity.

\end{abstract}


\section{Introduction}

Circinus X-1 is a very unusual X-ray binary system discovered in 1971
\citep{Mar71}. It undergoes outbursts at X-ray \citep{Kal76}, infrared
\citep{Gla78,Gla94} and radio \citep{Whe77,Hay78} wavelengths with a
period of 16.6 days, a fact that was interpreted as enhanced accretion
near the periastron of an elliptical orbit
\citep*{Mur80,Nic80}. Millisecond variability in X-ray led initially
to the classification of Cir X-1 as a black hole candidate
\citep{Too77}, but discovery of type I X-ray bursts from the field of
Cir X-1 \citep*{Ten86a,Ten86b} suggested identification of the compact
object with a neutron star.  The quasi-periodic oscillations (QPOs)
and X-ray colors of the object on occasion exhibit behaviour similar
to that of Z sources \citep{Shi99} but at other times Cir X-1
resembles an atoll source \citep{Oos95} and the source does not fit
perfectly to either of these two flavours of low magnetic field
neutron stars (see \cite{Kli95}). Optical/IR observations point to a
low mass X-ray binary system (LMXB) \citep*{Joh99}. Recently,
\cite*{Bou06} found twin kHz QPOs in the power spectra offering more
evidence for associating the compact object in Cir X-1 with a neutron
star. The orbit of the binary is not known but probably has a high
eccentricity \citep{Mur80, Nic80,Tau99}. Cir X-1 lies within a
synchrotron nebula \citep{Hay86} and has variable radio flux densities
at cm wavelengths. While the radio flares reached up to 1 Jy in
the late 1970s \citep{Whe77,Hay78}, they have only been observed at
the tens of mJy level ever since (e.g.  \cite*{Ste91,Fen05}).  The age
of the system is unknown, almost all of the estimates being made
within the framework of the ``runaway binary'' scenario that arose
from the hypothetical association with the nearby supernova remnant
(SNR) G321.9-0.3 \citep*{Cla75}. However, more recent Hubble Space
Telescope (HST) observations have revealed no evidence for proper
motion of Cir X-1, almost certainly ruling out such a relation
\citep{Mig02}, leaving the age of Cir X-1 extremely poorly
constrained. Another important parameter, the distance, has a very
large range of possible values, ranging from 4 to 9 kpc
\citep{Gos77,Iar05}. This system also seems to harbour the most
relativistic outflow observed within our galaxy \citep{Fen04}.  The
arcsec scale jet has a minimum bulk Lorentz factor of around 10 and an
inclination with respect to the line of sight of less than $\sim 5$
degrees, depending on the accepted distance. As a result, Cir X-1
resembles, at least from a geometrical point of view, a down-scaled
version of a BL Lac object.

\section{Observations}

We have observed Cir X-1 over five epochs between October 2000 and
September 2004 at 1.4 and 2.5 GHz using the Australia Telescope
Compact Array (ATCA). For better uv coverage, four different antenna
configurations were employed, with baselines ranging from 31 up to
6000 m. Table 1 contains the observational log. The absolute flux
density calibration was scaled with respect to PKS J1939-6342 (PKS
B1934-638). For phase calibration the nearby PMN J1524-5903 (B1520-58)
was used. At each of the two frequencies the data were calibrated for
the five epochs individually with MIRIAD software \citep*{Sau95} and
then combined together into an image that covered this way the widest
range of baselines permitted by ATCA. In our study we also made use of
the publicly
available\footnote{http://www.astrop.physics.usyd.edu.au/SUMSS/index.html}
0.8 GHz Molonglo Observatory Synthesis Telescope (MOST) data (FITS
image from April 3, 1992) which is part of the Sydney University
Molonglo Sky Survey (SUMSS) \citep*{Boc99}.

The nearby projection on the sky of SNR G321.9-0.3 (see Fig. 1)
resulted in artifacts on the radio maps, degrading the quality of the
images (the ATCA full width at half power primary beam is 33 arcmin
and 22 arcmin at 1.4 and 2.5 GHz respectively). As a result of this
and the diffuse nature of the Cir X-1 nebula itself, the images and
flux measurements of the nebula vary significantly with the chosen
angular resolution (which we can select by adjusting the weighting on
different baselines). Furthermore, for the total
flux measurements used in section 2.2 we weighted the short baselines
most strongly in order to recover as much of the diffuse nebular
emission as possible (while still trying to exclude artifacts from SNR
G321.9-0.3).

\begin{table*}
 \centering
  \begin{minipage}{100mm}
  \caption{Observational log.}
  \begin{tabular}{@{}lccc}
  \hline
   Epoch & Time [h] & Array configuration\footnote{A list of standard ATCA array configurations at: \\ 
http://www.narrabri.atnf.csiro.au/operations/array\_configurations/configurations.html.} & Frequency [GHz] \\
\hline
October 1, 2000 & 9 & 6A & 1.4, 2.5 \\ 
October 14, 2000 & 11 & 6A & 1.4, 2.5 \\
October 25, 2000 & 12 & 6C & 1.4, 2.5 \\
August 3, 2001 & 11 & 1.5A & 1.4, 2.5 \\
September 3, 2004 & 10 & EW214 & 1.4, 2.4 \\
\hline
\end{tabular}
\end{minipage}
\end{table*}

\begin{figure}
  \includegraphics*[scale=0.55]{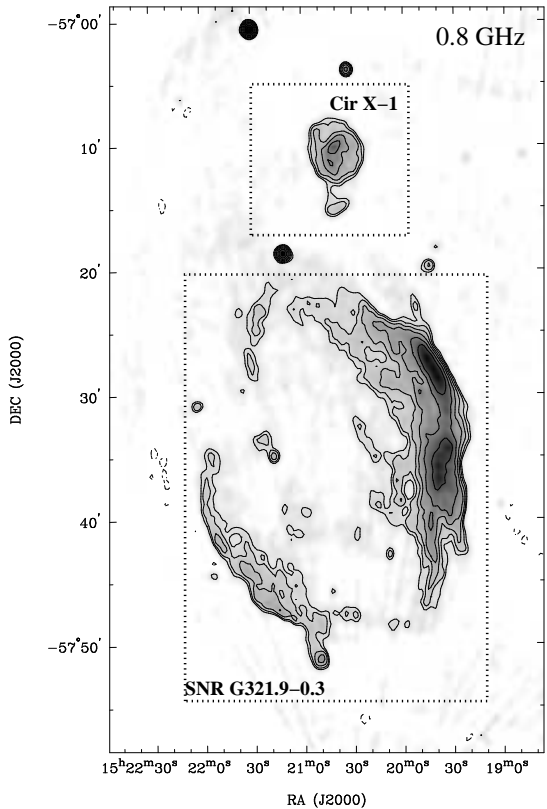}
  \caption{MOST total intensity 0.8 GHz map of the SNR G321.9-0.3 and radio nebula around Cir X-1. The contours are at -1, 1, 1.4, 2, 2.8, 4, 
5.6, 8, 11, 16, 23, 32, 45, 64 and 90 times the rms noise of 8.7 mJy/beam.}
\end{figure}

\begin{figure*}
  \includegraphics*[scale=0.6]{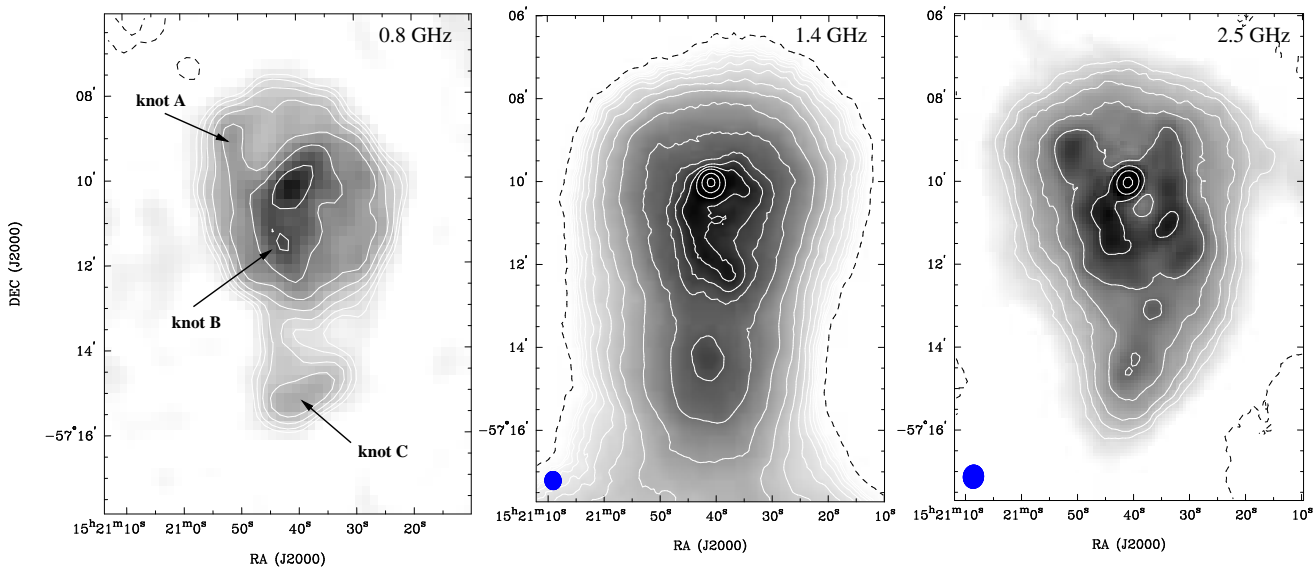}
  \caption{\textit{Left}: MOST total intensity 0.8 GHz map of the
    radio nebula around Cir X-1. The contours are at -0.5, 0.5, 0.7,
    1, 1.4, 2, 2.8, 4, 5.6, 8, 11, 16, 23, 32, 45, 64 and 90 times the
    rms noise of 8.7 mJy/beam.  The beam size is 51 $\times$ 43 arcsec$^2$ 
at PA=0$\degr$.0.  \textit{Center}: ATCA total intensity 1.4 GHz
    map of the radio nebula around Cir X-1. The contours are at -1, 1, 
    1.4, 2, 2.8, 4, 5.6, 8, 11, 16, 23, 28, 32, 37, 45, 64 and 90
    times the rms noise of 0.7 mJy/beam. The beam size is 26.2
    $\times$ 23.5 arcsec$^2$ at PA=4$\degr$.9. Larger than nominal
    ATCA beam size was used in order to show maximum of details at all
    scales.  \textit{Right}: ATCA total intensity 2.5 GHz map of the
    radio nebula around Cir X-1. The contours are at -0.7, 0.7, 1,
    1.4, 2, 2.8, 4, 5.6, 8, 11, 16, 23, 32, 45, 64 and 90 times the
    rms noise of 6.1 mJy/beam. The beam size is 33.7 $\times$ 29.4
    arcsec$^2$ at PA=-6$\degr$.8. Larger than nominal ATCA beam size
    was used in order to show maximum of details at all scales.}
\end{figure*}

\begin{figure*}
  \includegraphics*[scale=0.69]{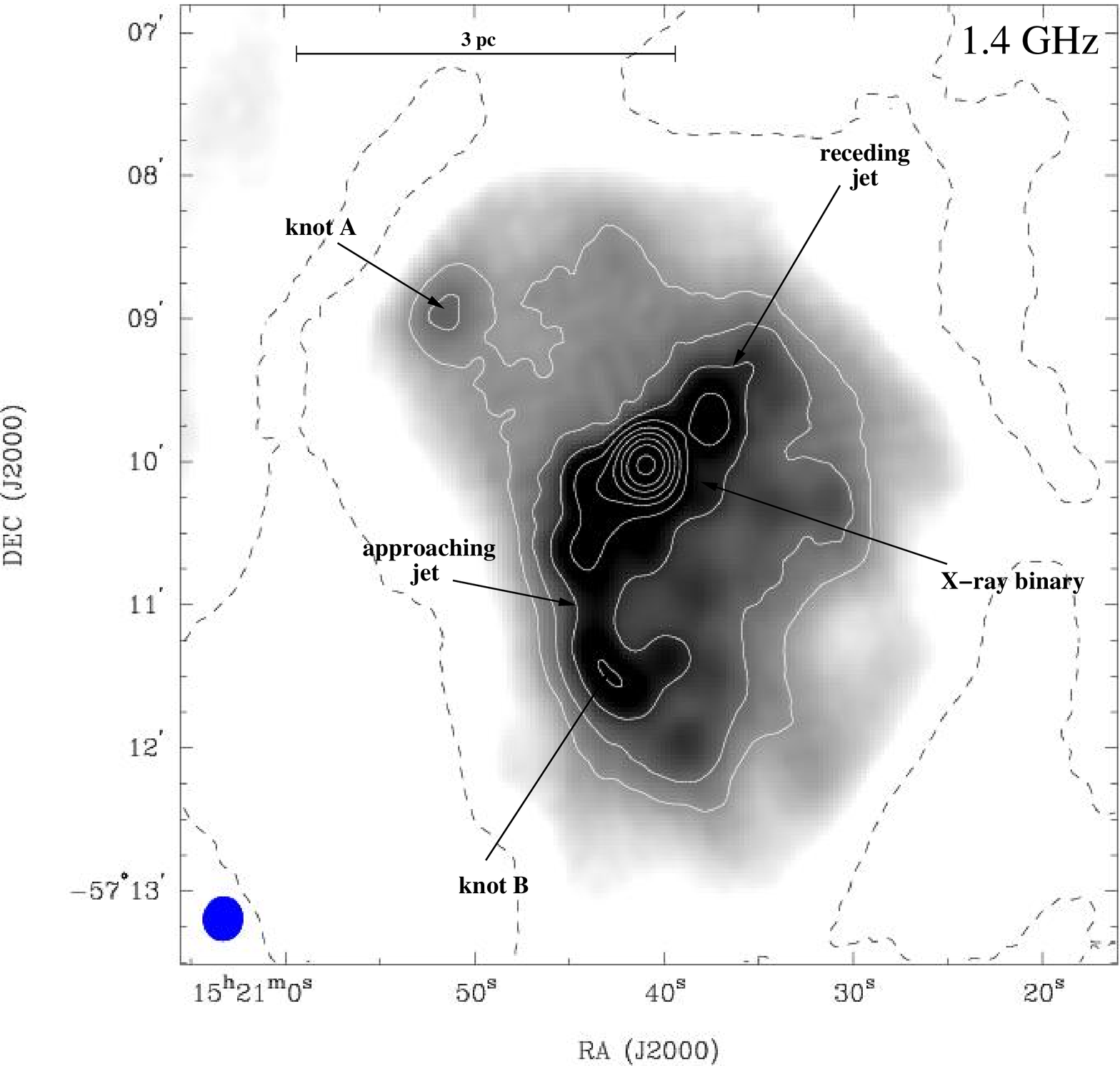}
  \caption{ATCA total intensity 1.4 GHz map of the central part of
radio nebula around Cir X-1 with data from August 2001 only (i.e. the
data set doesn't contain the shorter baselines from epoch September 3,
2004 and is therefore less sensitive to the diffuse radio emission,
neither the data from October 2000 which have a poorer u-v coverage
due to frequency switching). The contours are at -1, 1, 1.4, 2, 2.8,
4, 5.6, 8, 11, 16, 23, 32, 45, 64 and 90 times the rms noise of 2.0
mJy/beam.  The beam size is 18.1 $\times$ 16.5 arcsec$^2$ at
PA=-7$\degr$.1. Larger than nominal ATCA beam size was used in order
to show maximum of details at all scales.}
\end{figure*} 

\subsection{Radio maps}

In Fig. 1 we present the large-scale 0.8 GHz radio map of the Cir X-1
region from SUMSS, indicating the nearby SNR G321.9-0.3. Fig. 2 is a
montage of 0.8, 1.4 and 2.5 GHz maps of the nebula around Cir X-1,
clearly showing structure and emission extending in the direction of
the SNR, and finally Fig. 3 presents our most detailed 1.4 GHz image
of the nebula, revealing considerable structure within the
arcmin-scale jets.

An arcmin jet-like structure within the nebula (previously observed in
radio by \cite{Ste93}) is clearly seen at 0.8, 1.4 and 2.5 GHz in
Figs. 2 \& 3. The jet exhibits asymmetry: the south-eastern side (with
respect to the radio core) is slightly brighter and longer than the
north-western one.

As already noted, the arcsec scale jet of Cir X-1 may be the most
relativistic yet discovered in the galaxy, with an orientation very
close to the line of sight \citep{Fen04} and a position angle on the
sky coincident with the arcmin scale jet \citep{Fen98}. It therefore
seems reasonable to assume that the geometry of the jet is preserved
at arcmin scales, so the observed asymmetry at 1.4 GHz can be
explained in terms of an approaching (south-east) and a receding jet
(north-west) affected by projection effects. The bends of the jet may
be consequences of the precession affecting the disk of the X-ray
binary to which the jets are very likely coupled. Such precession of
the jets is observed in SS 433 (e.g. \cite{Mar79}), an X-ray binary
harbouring relativistic jets interacting with a surrounding radio
nebula, an object resembling Cir X-1.  Interactions with clouds of
higher density could be an alternative explanation for the change in
direction of the jet and this would appear to be observed at arcsec
scales at least \citep{Fen04}.

The expected ratio between the fluxes of the approaching and receding
continuous jets, for an inclination with respect to the line of sight
of 5-15 degrees and a velocity ratio $\beta=v/c= 0.05, 0.10, 0.15)$ is
(1.3, 1.6, 2.1). An image plane fit to the central region of the Cir
X-1 complex in the radio map in Fig. 3 (the inner 40 $\times$ 40
arcsec$^2$) shows that the simplest appropriate model is composed of
an unresolved source coinciding on the sky with the X-ray binary (at
the position given by Fender et al. 1998) and an extended source
centred some 2 arcsec to the south-east.  If we assume that the
SE-NW asymmetry is due to Doppler boosting, it follows that the radio
map at 1.4 GHz is consistent with a large-scale jet with a $\beta$
factor of at least 0.1. The maps at the other frequencies have a
poorer resolution and cannot add useful supplementary information in
this respect. As a caveat we should note that it is clear from the
results presented in Fender et al. (2004) that complex and varying
interactions do take place within the inner few arcsec, and there may
be a significant contribution to the SE-NW asymmetry from e.g. the
environment into which the jet is propagating.

An interesting feature present at all three frequencies is ``knot A''
in the north-east (Figs. 2 \& 3). Its nature is unknown.  It could be
a region of higher density that was energised by the passage of a
shock originating in the center of the complex. The blob is not
detected at higher frequencies (October, 2000 ATCA 4.8 and 8.6 GHz)
therefore an identification with a background radio source can
probably be ruled out, since the lack of detection at these
frequencies is likely due to a diffuse structure being resolved out.

An enhanced radio emitting region, ``knot B'', is evident at 0.8 and
1.4 GHz (Figs. 2 \& 3). This could be interpreted as the interaction
site where the approaching jet impinges on the interstellar
medium. However, projection effects might be at work and could
complicate such a simple explanation.

The ``knot C'' region (Fig. 2) also appears at all three frequencies
discussed here and looks rather isolated at 0.8 GHz and 1.4 GHz. The
nature of this is unknown. Its presence was easier to accommodate in
the ``runaway binary'' scenario, but the HST observations (see section
1) reopened the issue of its origin. It may be a relic of a previous
epoch of strong jet activity.

The likely general picture of the radio nebula emerging is that of a
system in which the energy and matter are supplied from the central
source via the jets. In this respect, we propose that the radio nebula
is formed at the interaction site between the jet, inclined at a low
angle to the line of sight, and the interstellar medium
(Fig. 4). Therefore, uniquely in the case of this system, we are
observing the core of the binary through one of the radio lobes.

\begin{figure}
  \includegraphics*[scale=0.35]{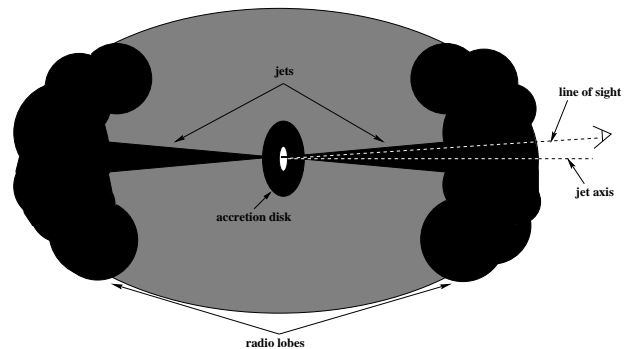}
  \caption{The proposed geometry of the Cir X-1 complex (not to scale)
  based on analogy with the W50/SS433 complex.}
\end{figure}

\subsection{Spectral index}

Even though the radio monitoring of Cir X-1 was sparse in the last
three decades, apparently there is a secular decrease in the radio
flux levels measured in the active as well as in the quiescent states,
with short periods of enhanced activity
\citep{Whe77,Hay78,Nic80,Pre83,Ste91,Ste93,Fen97,Fen98,Fen05}. The
ASM/Rossi X-ray Timing Explorer (RXTE)\footnote{http://xte.mit.edu/}
data from the last $\sim$ 10 yr show a similar behaviour in X-ray
band. Fig.5 presents the quiescent overall spectrum of the Cir X-1
complex (i.e. binary system plus radio nebula) from combined ATCA and
MOST data (see section 2). This is the first low-frequency spectrum
reported for this object since the late '70s \citep{Whe77,Hay78},
early '80s \citep{Hay86}. The spectral index obtained here, $\alpha
\sim -0.5 \ (F_{\nu} \propto \nu^{\alpha})$, is within the errors
identical to earlier values, with the flux level scaled down by $\sim$
300 mJy at 1.4 GHz in comparison with the measurements (notably at a
slightly different resolution) made in 1977 \citep{Hay78}. This is
almost certainly due to the secular decline of the radio core and we
find no evidence for a change in the radio luminosity of the nebula.

The spectral index map of the radio nebula, made with the combined
ATCA data at 1.4 and 2.5 GHz, is presented in Fig. 6. A larger than
nominal beam was used in order to increase the signal to noise ratio
of the extended emission (nevertheless, due to the simultaneity of the
observations at the two frequencies, we need to bear in mind that the
u-v coverage is not identical). Maps made with better resolution,
which at least in principle should be qualitatively correct, are
compatible with the features seen in Fig.6. Namely, there is a
tendency towards a flatter spectrum in the north-eastern part of the
nebula in a zone coincident with an enhanced radio emission region
(``knot A'' in Figs. 2 \& 3). This could be a site of augmented
electron acceleration.  Also, at a lower confidence level, the
spectrum seems to flatten at the edges of the nebula. If real, this
could be interpreted in terms of free-free emission or synchrotron
radiation from enhanced accelerated particles at the interaction site
with the interstellar medium.  The ``knot C'' (Fig. 2) doesn't appear
on the map due to our chosen cut-off in the confidence level of the
features represented.

\begin{figure}
  \includegraphics*[angle=-90,scale=0.34]{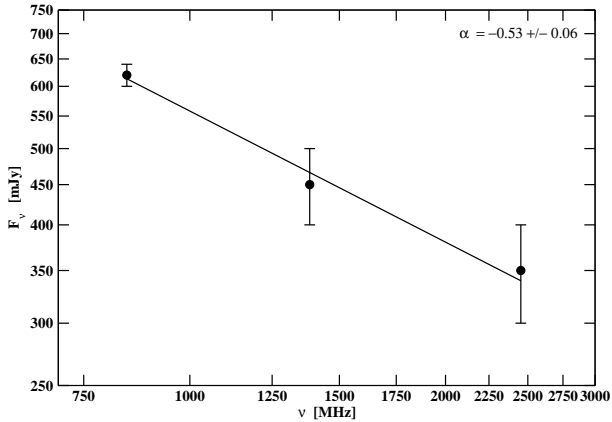}
  \caption{The overall spectrum of Cir X-1 complex in the quiescent state.}
\end{figure}

\begin{figure}
  \includegraphics*[angle=-90,scale=0.4]{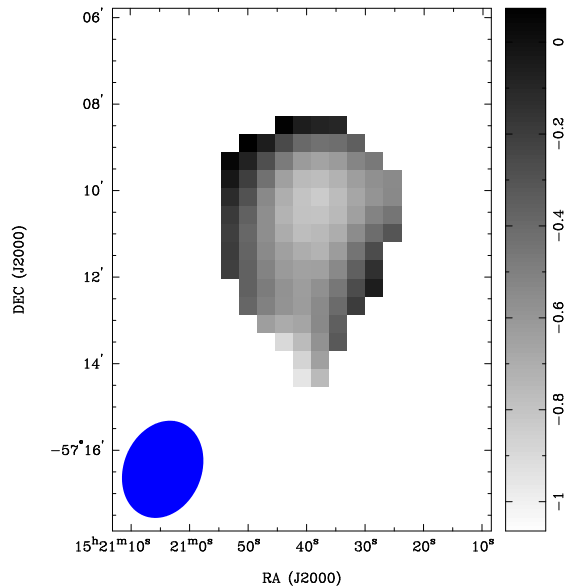}
  \caption{The spectral index map of the radio nebula around Cir X-1 between 1.4 and 2.5 GHz. The beam size is 139 $\times$ 108 arcsec$^2$ at 
PA=-22$\degr$.4}
\end{figure}

\section{Energetics}

\subsection{Minimum energy}

Assuming a source of volume $V$, specific luminosity $L_{\rm{\nu}}$, spectrum of the form $L_{\rm{\nu}} \propto \nu^{\alpha}$ and 
considering 
the radiation to be of synchrotron origin produced by a particle population with the energy spectrum $N(E) dE = N_{\rm{0}} E^{-p} dE$, 
it is possible to estimate the minimum energy associated with the source. The spectral index $\alpha$ is related to the particle 
distribution index $p$ by $p=1-2 \alpha$,  $N(E) dE$ is 
the number density of particles in the interval $[E, E+dE]$ and $N_{\rm{0}}$ is a proportionality constant. Two main pathways have been 
followed in literature for determining the relevant formulas: one using a fixed interval in 
frequency in the radiation spectrum \citep{Bur56,Pac70,Lon94}, the other in the particle energy spectrum(e.g. 
\citealt*{Poh93,Bru97,Pfr04}). We will adopt the former method and closely follow \citet{Lon94}. 

The internal energy associated with the observed synchrotron emission can be split in two components: $E_e$ the energy of 
particles (i.e. electrons and protons) and $E_B$ the energy of fields (i.e. magnetic):

\begin{equation}
E_{\rm{e}}=C_{\rm{e}} (1+\eta) L_{\rm{\nu}} B^{-3/2},
\end{equation}

\begin{equation}
E_{\rm{B}}=C_{\rm{B}} f V B^{2},
\end{equation}
where $B$ is the magnetic flux density, $\eta$ is the ratio between the energies in protons and electrons, $f$ is a parameter taking 
into account the uncertainties in the estimation of the volume $V$ of the emitting region (it is not what usually is referred to as the 
volume filling factor because we allowed it to be higher than 1, covering this way also the situations when an underestimation of the volume 
is made; this is very relevant in the case of Cir X-1 because there may be large projection effects along the line of sight). 
$C_B$ is a numerical constant related to the magnetic constant $\mu_0$ , while $C_e$ incorporates the dependences on the radiation spectrum 
and particle energy distribution (e.g. \cite{Lon94}).

The sum of energies in eqs. (1) and (2) has a minimum with respect to the magnetic flux density at :
\begin{eqnarray}
B_{\rm{min}} &=& 2.31 \times 10^{-2} \ C_{\rm{e}}^{2/7} (1+\eta)^{2/7} f^{-2/7} \left(\frac{V}{\rm{m}^3} \right)^{-2/7} \nonumber \\ 
&\times& \left(\frac{L_{\nu}}{\rm{W Hz^{-1}}} \right)^{2/7} \qquad \mbox{T},
\end{eqnarray}
thus the minimum energy necessary to explain the observed synchrotron emission reads:
\begin{eqnarray}
E_{\rm{min}} &=& 4.97 \times 10^{2} \ C_{\rm{e}}^{4/7} (1+\eta)^{4/7} f^{3/7} \left(\frac{V}{\rm{m}^3} \right)^{3/7} \nonumber \\ 
&\times& \left(\frac{L_{\rm{\nu}}}{\rm{W Hz^{-1}}} \right)^{4/7} \qquad \mbox{J}.
\end{eqnarray}
$B_{\rm{min}}$ and $E_{\rm{min}}$ correspond to a particular state of the system in which the energy in magnetic field is three 
quarters of the energy in particles.

Averaging the energy loss rate by synchrotron radiation over an isotropic distribution of pitch angles, the 
lifetime of an electron emitting at the peak of the synchrotron spectrum is:
\begin{eqnarray}
t &=& 8.5 \times 10^5 \left(\frac{\nu}{\rm{Hz}} \right)^{-1/2} \left(\frac{B_{\rm{min}}}{\rm{T}} \right)^{-3/2} \qquad \mbox{s}.
\end{eqnarray}

\subsection{Application to Cir X-1}

In order to estimate the minimum energy requirements for Cir X-1 we use eq. (4). The nebula was detected between 
0.4 GHz (\cite{Whe77}; considering that at low frequencies the radiation from the diffuse nebula dominates) and 2.5 GHz (this work), 
therefore we make the assumption that within this frequency interval the spectrum 
is of synchrotron origin and has the slope derived in section 2.2, $\alpha=-0.5$. The energy $E_e$ associated with the particles is dominated 
by the lowest energy particles and so the above upper limit in the frequency domain is 
not essential in the calculation. The volume of the emitting region can be very roughly estimated in this particular case by assuming 
the radio lobe has a spherical geometry: for 
a 250 arcsec radius object at 4.1 kpc \citep{Iar05} the resulting volume (both radio lobes taken into account) is 
$2.2 \times 10^{52} \mbox{ m}^{3}$. We further assume 
that the synchrotron radiative processes are dominated by electrons and therefore the ratio between the energy in protons and electrons 
$\eta=0$. Applying eq. (4), the minimum energy (in the comoving frame) required to account for the observed radio emission, under 
the assumptions made above is $8.1 \times 10^{46} \mbox{ erg}$. This value is of the same order of magnitude as the one inferred for 
the W50 nebula around SS433 \citep{Dub98,Mol05}. This raises the possibility that Cir X-1 might not be a jet-powered radio nebula only, 
but maybe a distorted SNR as suggested for W50 (e.g. \cite{Gre04}). The corresponding 
magnetic field derived from eq. (3) is $6.3 \ \umu \mbox{G}$. The electron lifetime (eq. (5)) at 2.5 GHz is then $3.4 \times 10^7 
\mbox{ yr}$. 

This estimate does not include energy stored in bulk motions or in heating of protons and so could be a significant underestimate. 
Errors in the evaluation of the volume ($f\in[10^{-2},10^{2}]$) would modify the minimum energy by not more than a factor 10.

Assuming the energy in the nebula has not been injected continuously
(i.e. along the full orbit of the X-ray binary), Fig.7 shows the
dependence of the averaged jet power as a function of the age of
nebula and the duty cycle of the binary system. We investigate this
because the periodic X-ray flares correlate with the radio flares
\citep{Whe77,Hay78,Tho78,Fen97} suggesting enhanced accretion onto the
compact object and ejection of matter via the jets. It is realistic to
consider that the injection of energy from the binary to the radio
nebula is occurring mainly during the flare states and to identify the
duty cycle in Fig. 7 with the duration of the flares. The solid
(lowest) line in Fig. 7 corresponds to a duty cycle of unity. For an
active jet phase of less than one day (corresponding to a duty cycle
$\leq 0.06$), and an age of the nebula of $10^{4}-10^{5} \mbox{ yr}$ ,
the minimum jet power would be of the order of $10^{36}-10^{37} \mbox{
erg s$^{-1}$}$.

Fig. 8 presents the ASM/RXTE light curve of the central X-ray source
of the Cir X-1 complex. Over the period of the X-ray data plotted, the
averaged X-ray power (for a distance of 4.1 kpc) is $4.3 \times
10^{37} \mbox{ erg s$^{-1}$}$. Such a high power is typical for most
of the Z sources, which are accreting at a large fraction of the
Eddington limit.

Making the simple assumption that in the flare state the X-ray light
curve has a square shape and the increase in amplitude from the
quiescent level is of 50 ASM counts s$^{-1}$ (a more or less typical
value in the last few years), then the additional power in X-rays
during the flare is of the order $3.4 \times 10^{37} \mbox{ erg
s$^{-1}$}$.  In any case, these estimates suggest that the jet power
during flaring phases may be a significant fraction of the X-ray
power.

\begin{figure}
  \includegraphics*[angle=-90,scale=0.34]{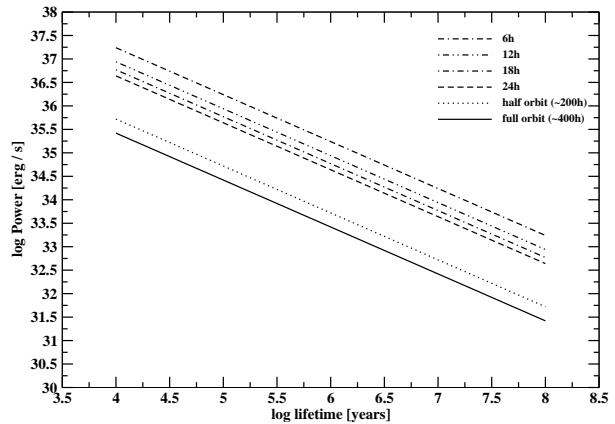}
  \caption{The averaged jet power of the radio nebula around Cir X-1 as a function of its age and the duty cycle of the 
binary system.}
\end{figure}

\begin{figure}
  \includegraphics*[angle=-90,scale=0.34]{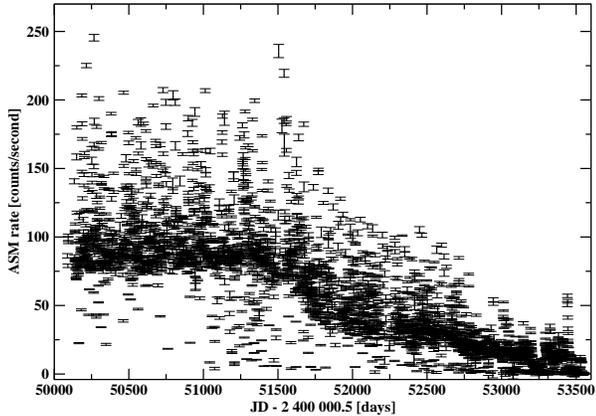}
  \caption{2-10 kev ASM/RXTE X-ray light curve of Cir X-1 between February 1996 - July 2005. }
\end{figure}

\section{Age of the radio nebula}

Constraining the age of the nebula is a very difficult task given the fact that so little is known about the Cir X-1 complex. Only very 
weak constraints can be imposed. 

The size of the nebula can offer a rough estimation of the age. Oversimplifying the problem and considering a constant expansion of 
the nebula with an average velocity in the plane of the sky $v$, the time required to achieve its present size (assuming a distance 
of 4.1 kpc) is:

\begin{equation}
t_{\rm{expansion}} = 2.0 \times 10^{4} \left(\frac{\theta}{\rm{arcsec}} \right) \left(\frac{v}{\rm{km \ s^{-1}}} \right)^{-1} \qquad \mbox{yr},
\end{equation}
where $\theta$ is the angular size. For $v=(10; 50; 100; 500; 1000 \mbox{ km s$^{-1}$})$, the expansion time is $t_{\rm{expansion}}=
(4 \times 10^{5}; 8 \times 10^{4}; 4 \times 10^{4}; 8 \times 10^{3}; 4 \times 10^{3} \mbox{ yr})$. Therefore if the nebula is 
expanding with a velocity comparable to that of the jet powered nebula around Cyg X-1 \citep{Gal05}, that is tens up to a few hundreds 
$\mbox{ km s$^{-1}$}$, its age is likely to be less than $\sim 10^5 \mbox{ yr}$. 

\section{A self-similar fluid model}

In order to attempt to understand the energetics and evolution of the
nebula better, we have developed a simple self-similar fluid model for
the jet-powered nebula.

We identify the radio nebula of Cir X-1 with the radio synchrotron
lobes inflated by the jets. This situation is analogous to the model
for the lobes of radio galaxies described by \citet{Kai97}. In the
following we develop a simplified version of this model for Cir
X-1. The main assumptions of the model are: the jets are in pressure
equilibrium with the lobes they inflate. The energy transport rate of
a single jet or ``jet power'', $Q_0$, is constant over the lifetime of
the jet. The jet may go through phases of outbursts and quiescence as
long as the duration of these phases is short compared with the
overall jet lifetime. In this case, $Q_0$ is the jet power averaged
over the lifetime of the jet.  The expansion of the lobe is confined
by the ram-pressure of the receding external gas in all
directions. Under these conditions \citet{Kai97} showed that the
expansion of the lobe is self-similar, e.g. the ratio $R=L/\left( 2 r
\right)$, where $L$ is the length of the lobe in the direction of jet
propagation and $r$ is the radius of the lobe perpendicular to the jet
measured at a fixed fraction of $L$, say $L/2$, is a constant.

\begin{figure}
\centerline{
\includegraphics[width=8.45cm]{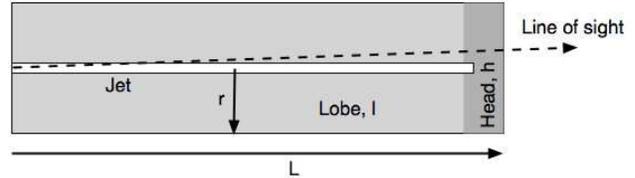}}
\caption{Cylindrical radio lobe seen from the side. In Cir X-1 our line of sight is very close to the jet axis. The relative size of the 
head region is exaggerated.}
\label{schem}
\end{figure}

We now simplify the model by assuming a cylindrical geometry for the lobe with the axis of the cylinder aligned with the jet. Figure 
\ref{schem} shows a schematic of the lobe geometry. The jet ends in a high-pressure region, the ``head'', where the energy transported by the 
jet is thermalized. The pressure in the head, $p_{\rm h}$, mainly drives the forward expansion of the lobe, which is confined by the 
ram-pressure of the external medium of mass density $\rho$. This can be expressed as:
\begin{equation}
p_{\rm h} \approx \dot{L}^2 \rho,
\label{ram}
\end{equation}
where we assume that the density of the external medium, $\rho$, is constant. The sideways expansion of the lobe is mainly driven by the 
pressure in the lobe, $p_{\rm l}$, and so
\begin{equation}
p_{\rm l} \approx \dot{r}^2 \rho.
\label{equ}
\end{equation}
The ratio of the two pressures is then:
\begin{equation}
\frac{p_{\rm h}}{p_{\rm l}} \approx \left( \frac{\dot{L}}{\dot{r}} \right)^2 = 4 R^2,
\end{equation}
where we have used the condition of self-similar expansion, i.e. $r= L / \left( 2 R \right)$ with $R = {\rm constant}$.

The volume of the head region, $V_{\rm h}$, will usually be small compared to the volume of lobe, $V_{\rm l}$ and so $V_{\rm h} / V_{\rm l} 
= R_V \ll 1$ \citep{Kai99}. The self-similar expansion of the lobe suggests that $R_V = {\rm constant}$, but we will see in the following 
that as long as $R_V \ll 1 / \left( 4 R^2 \right)$ any time dependence of $R_V$ is unimportant.

The total internal energy of the lobe, $U$, contained in $V_{\rm l}$ changes according to:
\begin{equation}
{\rm d} U = \frac{1}{\gamma - 1} \left( V_{\rm l} {\rm d} p_{\rm l} + p_{\rm l} {\rm d} V_{\rm l} \right) = Q_0 {\rm d} t - p_{\rm l} {\rm d} 
V_{\rm l} - p_{\rm h} {\rm d} V_{\rm h},
\label{balance}
\end{equation}
where $\gamma$ is the adiabatic index of the lobe material. The last term describes the expansion work of the head region. Using the results 
of the discussion above we find:
\begin{equation}
p_{\rm h} {\rm d} V_{\rm h} = 4 R^2 R_V p_{\rm l} {\rm d} V_{\rm l}.
\label{first}
\end{equation}
Thus we can eliminate all quantities referring to the head region from equation (\ref{balance}). From equation (\ref{equ}) and remembering 
that $r= L / \left( 2 R \right)$ we get:
\begin{equation}
{\rm d} t = \frac{1}{2R} \sqrt{\frac{\rho}{p_{\rm l}}} {\rm d} L.
\label{second}
\end{equation}
Finally, the self-similar expansion of the lobe and $R_V = {\rm constant}$ imply 
\begin{equation}
V_{\rm l} = V_0 \left( \frac{L}{L_0} \right)^3,
\label{third}
\end{equation}
where $L_0$ is an arbitrary scale length. 

Substituting equations (\ref{first}), (\ref{second}) and (\ref{third}) into equation (\ref{balance}) yields after re-arranging
\begin{equation}
\frac{{\rm d} p_{\rm l}}{{\rm d} L} = \frac{\gamma -1}{2 R} \frac{Q_0}{V_0} \sqrt{\frac{\rho}{p_{\rm l}}} \left( \frac{L_0}{L} \right)^3 -3 
\left[ \gamma + \left( \gamma - 1 \right) 4 R^2 R_V \right] \frac{p_{\rm l}}{L}.
\end{equation}
The solution of this equation is:
\begin{equation}
p_{\rm l} = p_0 \left( \frac{L}{L_0} \right)^{-4/3},
\end{equation}
with
\begin{equation}
p_0 = \left\{ \frac{\gamma - 1}{2 R \left\{ 3 \left[ \gamma + \left( \gamma - 1 \right) 4 R^2 R_V \right] - 4/3 \right\}} \frac{Q_0 L_0}{V_0} 
\sqrt{\rho} \right\}^{2/3}.
\end{equation}

From the observations we can determine $r$ and an estimate for $p_{\rm l}$. We then have to assume a value for the aspect ratio $R$, where 
$1 \le R \le 5$ is reasonable. With this we can determine the length of the lobe $L$. Since we are free to choose $L_0$, it is convenient to 
set $L_0 = L$. $V_0$ then follows from $L_0$ and $R$ and an assumption about $R_V$. If $R_V \ll 1 / \left( 4 R^2 \right)$, then $V_0 \approx 
\pi r^2 L_0 = \pi L_0^3 / \left( 4 R^2 \right)$ and we can neglect the term involving $R_V$ in  the eq.(21). With a typical value for 
$\rho$ appropriate for the external medium we can then determine $Q_0$, the time-averaged jet power.

Finally, the age of the jet/lobe system can be found from equation (\ref{second}) and is, for our choice $L_0 = L$, 
\begin{equation}
t = \frac{3}{10 R} \sqrt{\frac{\rho}{p_0}} L_0.
\end{equation}

We assume the lobe is populated by relativistic electrons and
therefore the adiabatic index $\gamma=4/3$. $p_h$ can be estimated
from the minimum energy requirements as discussed in section 3 and
using eq. (14) a minimum lobe pressure $p_l$ is obtained. For an
angular size of the lobe of 250 arcsec at 4.1 kpc, R=4, $R_V=0.01$ and
a number density of the external medium of 2 \mbox{cm$^{-3}$}, the
resulting long-term averaged jet power is $4 \times 10^{35} \mbox{ erg
s$^{-1}$}$ while the age is $2 \times 10^{4} \mbox{ yr}$.  For lower
duty cycles, as is very probably the case for Cir X-1, the jet power
estimated here could be up to two orders of magnitude higher.  The
orders of magnitude of the jet power and age of the nebula are weakly
dependent (i.e. do not affect the statements we make) on different
reasonable values assumed for the parameters of the model. For
instance, for a distance $d$ a factor two higher than the one we
preferred, the jet power becomes $7.5 \times 10^{34} \mbox{ erg
s$^{-1}$}$ and the age $1.1 \times 10^{5} \mbox{ yr}$.

With an averaged jet power of $\sim 10^{35} \mbox{ erg s$^{-1}$}$ and an age of the nebula of less than $\sim 10^5 \mbox{ yr}$, these 
results are compatible with the ones presented in section 3 (see Fig. 7). Clearly, it is not unphysical that the jet could have inflated the nebula.

\section{Conclusions}

We presented ATCA radio maps of the nebula around Cir X-1 at 1.4 and
2.5 GHz. Combining them with publicly available MOST data we obtained
the quiescent spectrum for the complex. The slope of the spectrum is
within the error identical to the ones obtained in the 1970's-1980's
but is scaled down by around 300 mJy at 1.4 GHz, compatible with the
observed secular evolution of the central source, both in radio and
X-rays. As a whole, the radio nebula can be seen as a galactic
analogue of the radio lobes in AGNs: the jet interacts with the
interstellar medium and creates the synchrotron radio
emission. Uniquely in this case, we are probably seeing the central
source through the jet-inflated radio lobe, as the jet appears to be
aligned very close to the line of sight.  In this respect, the Cir X-1
nebula may be an ``end-on'' analogue of the SS 433/W50 complex. The
behavior of the small scale jet \citep{Fen04} suggests that the energy
that powers up the nebula is supplied intermittently from the central
binary system via the jets. Even though hard to strongly constrain in
value, the amount of energy transferred to the nebula is important and
should be taken into account whenever an energy balance of the whole
system is required. The calculations suggest an averaged jet power of
at least $10^{35} \mbox{ erg s$^{-1}$}$ for a preferred age of the
nebula of $\leq$ 10$^5$ years and it may be that the jet power becomes
comparable to the X-ray luminosity during outbursts. Transfer of
considerable energy via jets is not rare, the case of SS433 being well
studied (e.g. \cite{Dub98}); moreover such a process might be more
common than previously suspected, as pointed out recently by
\cite{Gal05}.

Finally, we note that this is one of the best estimates to date for
the power in jets from a neutron star system. Such measurements may be
key in attempts to quantify the disc-jet coupling not only in other
neutron star, but also in black hole systems of all scales
\citep{Kor06}.

\section*{Acknowledgments}

We thank the anonymous referee for detailed comments. VT would like to
thank James Miller-Jones and Elena Gallo for useful discussions.  The
Australia Telescope is funded by the Commonwealth of Australia for
operation as a national facility managed by CSIRO.  The Molonglo
Observatory Synthesis Telescope (MOST) is operated by the School of
Physics of the University of Sydney. The X-ray data was provided by
the ASM/RXTE teams at MIT and at the RXTE SOF and GOF at NASA's GSFC.

\end{document}